\documentclass{vgtc}                          

\graphicspath{{figures/}{pictures/}{images/}{./}} 

\usepackage{times}                     

\usepackage{tabu}                      
\usepackage{booktabs}                  
\usepackage{lipsum}                    
\usepackage{mwe}                       

\usepackage{mathptmx}                  
\usepackage{balance}

\onlineid{0}

\vgtccategory{Research}

\vgtcinsertpkg

\title{Towards Embodied Conversational Agents for Reducing Oral Exam Anxiety in Extended Reality}

\author{Jens Grubert\thanks{e-mail: jens.grubert@hs-coburg.de}\\ %
        \scriptsize Coburg University of Applied Sciences and Arts %
\and Yvonne Sedelmaier\thanks{e-mail: yvonne.sedelmaier@hs-coburg.de}\\ %
     \scriptsize Coburg University of Applied Sciences and Arts 
\and Dieter Landes \thanks{e-mail: dieter.landes@hs-coburg.de}\\ %
     \parbox{1.4in}{\scriptsize \centering Coburg University of Applied Sciences and Arts}}

\abstract{
Oral examinations are a prevalent but psychologically demanding form of assessment in higher education. Many students experience intense anxiety, which can impair cognitive performance and hinder academic success. This position paper explores the potential of embodied conversational agents (ECAs) in extended reality (XR) environments to support students preparing for oral exams. We propose a system concept that integrates photorealistic ECAs with real-time capable large language models (LLMs) to enable psychologically safe, adaptive, and repeatable rehearsal of oral examination scenarios. We also discuss the potential benefits and challenges of such an envisioned system. 

} 

\keywords{mental health, embodied conversational agents, virtual agents, generative artificial intelligence, extended reality}

\begin{document}



\firstsection{Introduction}
\maketitle

Oral examinations are a critical but often overlooked source of academic stress in higher education \cite{stephenson2025interventions}. Unlike written assessments, oral exams require real-time verbal articulation of knowledge in socially charged and often unpredictable settings. Many students report intense nervousness, anxiety, and a lack of confidence when preparing for or undergoing oral exams. These affective barriers can severely impair cognitive performance—leading to poorer outcomes, diminished self-efficacy, and avoidance behaviors that undermine long-term learning.

Despite growing recognition of student mental health as a key concern in higher education, oral exam anxiety remains an underexplored area. Traditional support structures such as instructor-led mock exams or peer feedback are limited in availability, consistency, and personalization. Moreover, students may feel reluctant to expose their lack of preparedness or emotional vulnerability in front of real human evaluators.

Emerging technologies in immersive environments and conversational artificial intelligence offer new opportunities to support learners both cognitively and emotionally. Extended Reality (XR), with its variations of augmented, mixed, and virtual reality (AR/MR/VR), can simulate oral examination settings through embodied conversational agents (ECAs) with high fidelity, enabling students to repeatedly engage in realistic practice without the judgment of human observers. ECAs are computer-generated characters ``that demonstrate many of the same properties as humans in face-to-face conversation, including the ability to produce and respond to verbal and nonverbal communication" \cite{cassell2001embodied}. Recent advancements in Generative Artificial Intelligence (GenAI), specifically Large Language Models (LLMs), have led to a growing interest in combining ECAs with GenAI to create more intelligent behaviors (e.g., \cite{maslych2024takeaways, schmidt2020intelligent, schmidt2024natural,  sonlu2021conversational, yang2024effects, zhu2023free}).


In particular, we foresee that these systems could help students preparing for oral exams by offering psychologically safe rehearsal environments for repeated exposure to exam-like scenarios through real-time, adaptive feedback on both content and communication performance. Additionally, they could provide the opportunity to build self-confidence and manage exam-related anxiety through guided interaction and exposure-based learning.

The use of embodied agents in such contexts seems promising. By presenting a human-like presence with emotional expression, gaze behavior, and verbal turn-taking, ECAs could simulate examiner roles in a way that feels authentic and remains emotionally controllable for the learner. Crucially, these agents could be designed to support both knowledge rehearsal and emotional regulation. This could potentially reduce stress associated with oral assessment settings. However, realizing this potential requires addressing multiple challenges. 


This position paper outlines the potential of such systems, given a theoretical system concept, and discusses challenges that could arise when using ECAs in XR environments for exam preparation. 



\begin{figure*}[t!]
 \centering 
 \includegraphics[width=2\columnwidth]{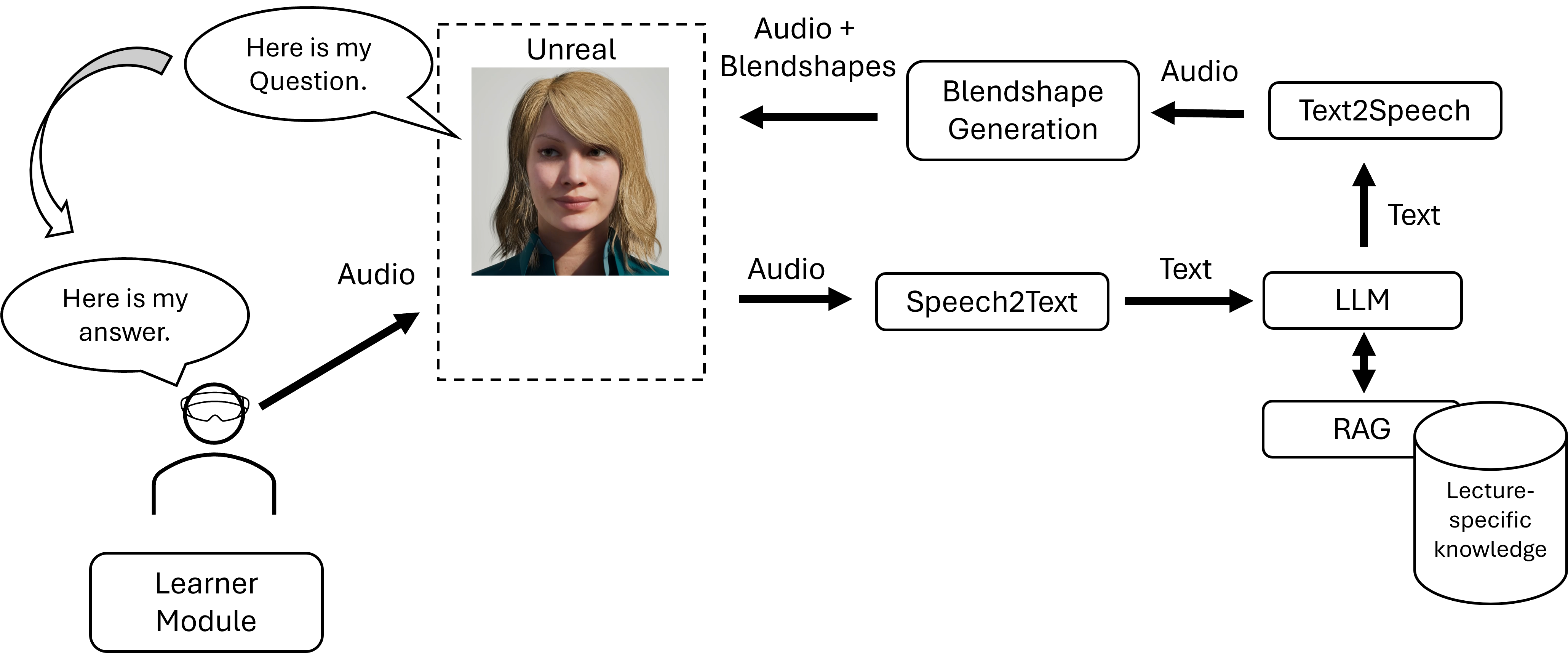}
 \caption{Envisioned technical pipeline for the interaction with the ECA in XR.}
 \label{fig:pipeline}
\end{figure*}

\section{Concept for ECAs for Oral Exam Preparation}

To address the emotional and cognitive challenges of oral exam preparation, we propose a system that integrates ECAs into XR learning environments (see also \autoref{fig:pipeline}). At its core, this system should function as an intelligent and domain-aware virtual examiner, enabling learners to repeatedly and independently rehearse oral exam scenarios in a psychologically safe yet realistic setting.
Our system leverages the recent convergence of various technological strands: 1) Photorealistic and behaviorally expressive virtual agents, which allow lifelike representations of human examiners. 2) Real-time natural language interfaces, enabled by LLMs combined with speech-to-text (STT) and text-to-speech (TTS) technologies, supporting fluid, unscripted dialogue between student and agent. 3) Domain-specific retrieval-augmented generation (RAG) pipelines to increase the chance for factual correctness and curricular alignment of the agent’s responses. 4) Adaptive, learner-centered feedback systems, capable of adjusting difficulty, tone, and guidance based on individual learner profiles and affective states.

We envision our system to consist of the following components: 
\paragraph{Embodied Agent} A virtual examiner rendered as a photorealistic avatar in a simulated exam room (in our case, using Unreal's MetaHuman platform). The agent displays naturalistic gaze, lip synchronization, facial expressions, and gestures, contributing to the sense of co-presence. The embodiment supports not only immersion but also models typical non-verbal cues that students must respond to during oral examinations.

\paragraph{Conversational Engine} Real-time dialogue is mediated by an LLM backend. Speech input from the student is transcribed via an STT module (e.g., using Whisper \cite{radford2023robust} or Voxtral \cite{liu2025voxtral}), processed for intent and content, and routed to the LLM (augmented by domain-specific retrieval). The output is synthesized via TTS, with synchronized facial animation (e.g., using NVIDIA's Audio2Face \cite{karras2017audio}) to produce the agent’s response.

\paragraph{Domain Knowledge Integration} To ensure content validity and alignment with the targeted academic domain (e.g., data mining), the system includes a curated knowledge base indexed for semantic retrieval. This includes lecture content, question pools, textbook excerpts, and annotated exam transcripts. The RAG layer enhances the LLM's responses by grounding them in authoritative curricular sources.

\paragraph{Learner Modeling and Feedback Module} This component maintains dynamic learner profiles, including cognitive markers (e.g., mastery level, prior responses), affective cues (e.g., speech hesitations, pacing), and usage data (e.g., repetition patterns, session duration). These models inform both the agent’s behavior (e.g., adaptively increasing challenge) and, optionally, post-session feedback reports.

\paragraph{Extended Reality Interface} The XR interface places the student in a virtual examination room. Complementary to the direct (speech and gaze-based) interaction with the ECA, we plan to include further information (e.g., as 2D overlays) to show supplementary, emotional feedback, or guidance during the session.

In the long term, we plan the virtual coach to function as part of a broader personalized learning ecosystem. Integration with digital learning platforms should allow the system to pull course-specific content, align with individual study plans, and provide cross-modal recommendations (e.g., suggesting a quiz after a weak verbal explanation of a concept). This integration also enables long-term tracking of progress and adaptation of feedback over time. Ideally, instructors should also be able to configure virtual “exam blueprints” that define question sets, difficulty levels, and evaluation rubrics.

\section{Potentials}

The psychological demands of oral examinations are not limited to content mastery—they also involve social presence, real-time judgment, and performance under pressure. These stressors can disproportionately affect students who otherwise possess the necessary knowledge, resulting in underperformance due to emotional overload rather than intellectual shortcomings. Within this context, our proposed ECA-based coach system could offer mental health support by intervening on three interconnected levels: exposure regulation, emotional scaffolding, and self-efficacy development.

One of the most robust psychological techniques for reducing anxiety is systematic desensitization through repeated, controlled exposure \cite{bruno1973systematic}. By simulating oral exam scenarios with increasing levels of fidelity, ECAs could allow students to become familiar with the format, rhythm, and social dynamics of oral assessments. As this exposure would occur in a low-stakes environment, anticipatory anxiety could be reduced, and learners could be supported in building tolerance and composure over time.

Unlike human-led mock exams, the ECA can be engaged at any time, with unlimited frequency, and without fear of judgment or embarrassment. The psychological safety of the system is (ideally) reinforced by its non-judgmental nature, enabling students to experiment, make mistakes, and retry challenging scenarios as often as needed. Over time, this repeated exposure can diminish the novelty and threat associated with oral examinations, replacing fear with routine and predictability.

Beyond mere exposure, the conversational agent can actively support emotional regulation during the learning process through emotional scaffolding \cite{liu2024improving}. By detecting stress signals, e.g. through speech disfluencies, long pauses, or increased vocal pitch, the system could respond in emotionally intelligent ways. For instance, the agent may (e.g., by adjusting the tone of responses, offering affirmations or breaks). This form of affective scaffolding mirrors techniques used by human tutors who are sensitive to student stress.

These interactions can normalize emotional reactions to challenging tasks, helping students to recognize, name, and manage their anxiety. The process fosters emotional literacy, which has been shown to correlate with academic persistence and mental health resilience.

Further, structured agent-led rehearsal and reflection can boost self-efficacy \cite{pareto2011teachable}. For example, when students are able to coherently explain a complex concept to the agent, or successfully handle a difficult follow-up question, they could receive immediate confirmation of their competence. Over time, these micro-successes can contribute to a heightened sense of capability and agency. 


Furthermore, the envisioned post-session feedback system could highlight not only what was incorrect but also what was handled well. Through this it could emphasize growth rather than deficiency. Visualization of progress over multiple sessions (e.g., increased fluency, better answer structure, more confident tone) could provide students with feedback on self-development, reinforcing a positive self-concept.


Finally, in diverse student populations, oral exams may also trigger identity-based vulnerabilities such as language insecurity among non-native speakers, fear of negative stereotyping, or prior experiences of academic invalidation. A well-designed ECA could act as a neutral, inclusive, and controllable partner that provides a sense of autonomy and equality not always found in real-life human interactions. This inclusivity could be reinforced by customizable features (e.g., agent gender, voice, demeanor). In fact, prior work has indicated the tendency of students to design ECAs that tend to resemble their own demographics \cite{feijoo2024exploring}.

Overall, by combining repeated exposure, emotional scaffolding, and self-efficacy building in a safe and controllable setting, ECA-based rehearsal could play a meaningful role in mitigating oral exam anxiety, thereby supporting both academic performance and long-term mental well-being.

\section{Challenges}
While the potential benefits of mixed reality coaching systems with ECAs are compelling, their development and deployment involve a series of challenges. 

For example, a preliminary internal survey at our university (n=44) \cite{Sedelmaier2025ICSEA} indicated that students consistently express the desire for authentic exam simulations including realistic examiner behavior, lifelike avatars, and faithful recreations of exam room environments, and this high realism supports immersion and transfer to real-life settings. However, for anxious students, overly realistic scenarios could potentially be emotionally overwhelming, especially if not scaffolded appropriately. This raises the question: How much realism is pedagogically optimal in a given learning setting? This has partly been addressed, e.g., with respect to the effects of behavioural and visual realism on knowledge gain \cite{petersen2021pedagogical} or arousal \cite{vicneas2019different}. A highly authentic simulation might mirror stress-inducing conditions too closely, potentially reinforcing avoidance or panic. On the other hand, low-fidelity environments risk undermining credibility and engagement. Hence, we foresee adjustable realism settings, allowing users to modulate factors such as the agent’s demeanor, environmental detail, and session pacing based on their comfort level and training stage. Ideally, this could be done in an automated and personalized way such that emotional intensity and realism increase over time, analogous to scaffolding techniques. 

Conversational dynamics in oral exams can be shaped not just by verbal content but also by subtle paralinguistic and nonverbal signals—such as interruptions, hesitation management, eye contact, facial expressions, and tone modulation. Simulating these aspects convincingly is an ongoing challenge for current ECAs. Also, LLM-based agents can still struggle with fine-grained turn-taking in spontaneous, spoken dialogue \cite{castillo2025survey}. Response latency or inappropriate phrasing can quickly disrupt the illusion of a natural interaction, but might be these effects might be mitigated by behavioural and verbal fillers of the ECAs \cite{Gonzales2025TVCG}. Moreover, learners likely notice and react to these breakdowns, especially in an emotionally loaded context. This, in turn, could result in reduced engagement or even frustration. 


While adaptive LLM-based systems promise individualized learning trajectories \cite{cordova2025ai}, personalization should be carefully designed to avoid cognitive or emotional overload. If the agent adapts too quickly, students may feel disoriented or pressured. If it adapts too little, the experience may feel generic and demotivating. Content difficulty, agent behaviour, feedback style, and session structure all need to be carefully balanced to support students in their preparation efforts. But even if this balance could be found, there might be a risk of opaque adaptivity, i.e. users being unsure why the system behaves in certain ways in one situation over the next. This, in turn, could lead to confusion or mistrust. It remains an open question how transparent personalization, e.g., via explainable AI strategies (“I’m asking this follow-up because you struggled with a similar concept last time”) might impact user agency and acceptance.






Finally, unlike many educational technologies, where users might tolerate imperfection in early versions, emotionally sensitive systems such as exam preparation coaches likely operate under low tolerance for initial failure. Students preparing for high-stakes oral assessments might be unlikely to engage with a system that feels unreliable, unempathetic, or unfinished—especially when their stress levels are high. This could lead to a first impression bottleneck. If the first interaction feels artificial or inaccurate, users may abandon the system permanently, regardless of its long-term potential.

Addressing these challenges is crucial not only for ensuring the technical and pedagogical effectiveness of such systems but also for safeguarding students’ psychological safety and reducing the mental health burden often associated with oral examinations.


\section{Conclusion}

In this paper, we outlined the use of ECAs in XR to support oral exam preparation. We argued that ECAs can offer students psychologically safe, adaptive, and  realistic rehearsal experiences. By simulating the cognitive and emotional demands of oral assessments, such systems may help reduce anxiety, strengthen self-efficacy, and improve oral performance over time. At the same time, several challenges remain that should be addressed to ensure user trust, learning effectiveness, and emotional safety. To this end, we call for interdisciplinary collaboration between researchers in AI, education, HCI, and mental health to further investigate the impact of ECAs on learner well-being and learning success.  


\balance
\bibliographystyle{abbrv-doi}

\bibliography{template}
\end{document}